\documentclass[review,number,sort&compress]{elsarticle}

\usepackage{siunitx}
\usepackage{amssymb}
\usepackage{subcaption}

\journal{Nuclear Instruments and Methods A}

\begin{document}


\begin{frontmatter}
\title{The ultimate performance of the Rasnik 3-point alignment system}

\author[nikhef,delft]{Harry van der Graaf\corref{mycorrespondingauthor}}
\cortext[mycorrespondingauthor]{Corresponding author at: Nikhef, Science Park 105, 1098 XG Amsterdam, The Netherlands; retired}
\ead{vdgraaf@nikhef.nl}

\author[nikhef]{Alessandro Bertolini}
\author[louvain]{Joris van Heijningen}
\author[asi]{Bram Bouwens}
\author[amolf]{Nelson de Gaay Fortman}
\author[uva]{Tom van der Reep} 
\author[g]{Lennart Otemann}

\address[nikhef]{Nikhef, Science Park $105$, Amsterdam, The Netherlands}
\address[delft]{Delft University of Technology, Mekelweg $1$, Delft, The Netherlands}
\address[louvain]{Centre for Cosmology, Particle Physics and Phenomenology (CP3), Universit\'{e} catholique de Louvain, Louvain-la-Neuve, Belgium}
\address[asi]{Amsterdam Scientific Instruments, Science Park 106, Amsterdam, The Netherlands}
\address[amolf]{AMOLF, Science Park $104$, Amsterdam, The Netherlands}
\address[uva]{now at University of Amsterdam, Nieuwe Achtergracht $129$, Amsterdam, The Netherlands}
\address[g]{master student at Nikhef and TU Delft}

\begin{abstract}
	The Rasnik system is a $3$-point optical displacement monitor with sub-nanometer precision. The CCD-Rasnik alignment system was developed in $1993$ for monitoring the alignment of the muon chambers of the ATLAS Muon Spectrometer at CERN. Since then, the development has continued as new CMOS imaging pixel chips became available. In this work the system processes and parameters that limit the precision are studied. We conclude that the spatial resolution of Rasnik is only limited by the quantum fluctuations of the photon flux arriving at the pixels of the image sensor. The results of two Rasnik systems are compared to results from simulations, which are in good agreement. The best spatial resolution obtained was $\SI{7}{pm/\sqrt{Hz}}$. Finally, some applications of high-precision Rasnik systems are set out.
\end{abstract}


\end{frontmatter}

\section{Introduction}\label{sec:intro}
The \textbf{R}ed \textbf{A}lignment \textbf{S}ystem \textbf{Nik}hef (Rasnik) was originally developed for the alignment of the inner, middle and outer muon chambers of the Muon Spectrometer of the L$3$ experiment at CERN ($1980$ - $2001$)~\cite{beker_2019,patent}. Units of three position-sensitive detectors, placed at a certain mutual distance, were used to measure the radius-of-curvature of muon tracks, from which the muon momentum can be derived. The precision of the measured muon momentum depends crucially on the alignment of the three chambers. Since it is practically impossible to mechanically constrain the three chambers in their nominally aligned position, the actual alignment is monitored continuously instead. The measured radius-of-curvature is corrected using the data from the alignment system. In this way, the muon momentum measurement is no longer affected by any change in the position of a muon tracking chamber.

Data from the 8000 operational ATLAS Rasnik systems revealed a much better intrinsic performance than the original ATLAS requirement (30 $\rm{\mu}$m).
In several studies, a sub-$\rm{\mu}$m spatial resolution was demonstrated~\cite{beker_2019}.
This paper reports the results of MonteCarlo simulations, and the results of two CCD-Rasnik systems, build for this purpose.
First, in section~\ref{principles}, the basic principles of the CCD-Rasnik system are set out.
In  section~\ref{ErrorSources}, all error sources are addressed and their contribution is analysed and quantified. Then, in section~\ref{analysis}, the MonteCarlo (MC) image generator is set out, followed by a description of the image analysis routine. By using this routine for analysing MC images, the performance of Rasnik systems can be estimated. The expected performance is compared to the results from the two hardware Rasnik systems in section~\ref{Results}. Finally, possible improvements for future systems are listed and quantified.

\section{Principles}\label{principles}
\begin{figure}[htb]
\centering
\includegraphics[width=\textwidth]{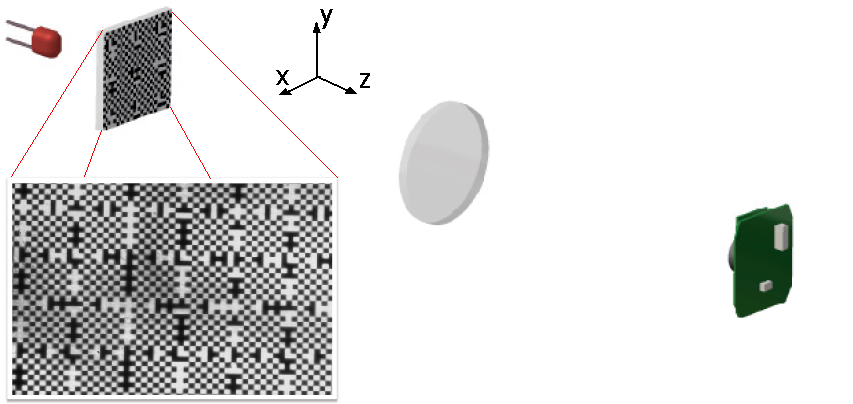}
\caption{Principle of the CCD-Rasnik system. An image of the LED back-illuminated coded mask is projected onto the image pixel sensor by means of a positive singlet lens. The sensor is connected to a PC by means of Ethernet or USB. The images are processed, resulting in four output parameters describing the alignment of mask, lens and sensor. If the lens and image sensor are fixed together (forming a camera), Rasnik turns into a system that monitors the $4$D position of the coded mask. This figure is reproduced from ref~\cite{beker_2019}.}
\label{fig:Principle}
\end{figure}

The basic principle of the CCD-Rasnik alignment system is shown in fig.\ref{fig:Principle}. The \textbf{optical axis}, running from the centre of the image sensor through the optical centre of the lens, crosses the mask at a specific point. This point is projected in the centre of the images from the sensor. The image analysis routine yields an $x$ and $y$ coordinate of this crossing point, with the mask pattern defining the coordinate system. Given the known mask pattern and the pitch of the pixels of the image sensor, the scale of the image can be obtained as third output parameter. This scale S equals the ratio of the image and object distances: S = $1$ if the lens is in the centre between the mask and the sensor. The fourth output parameter is the differential rotation $\theta_z$ of the mask and the sensor around the optical ($z$) axis.

\begin{figure}[htbp]
	\centering
	\begin{subfigure}[b]{0.45\textwidth}
		\includegraphics[width=\textwidth]{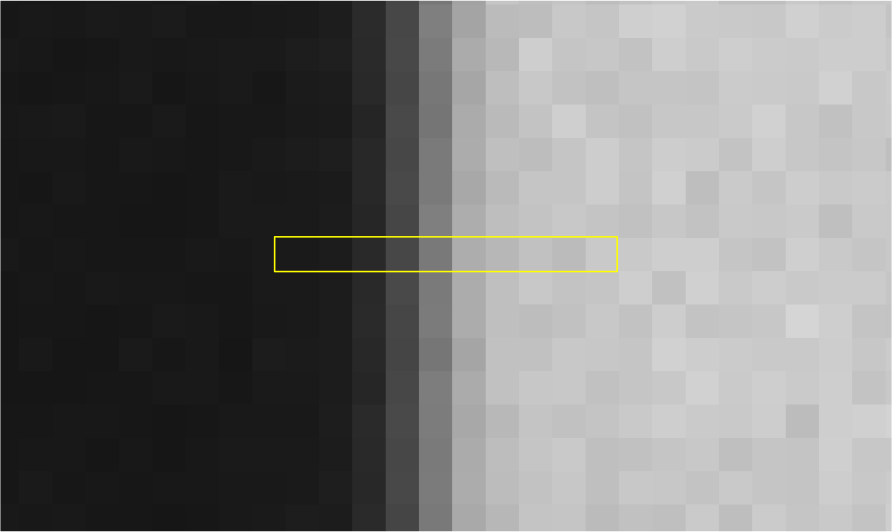}
		\vspace*{0.10cm}
		\caption{}\label{fig:pixcon}
	\end{subfigure}
	\begin{subfigure}[b]{0.49\textwidth}
		\includegraphics[width=\textwidth]{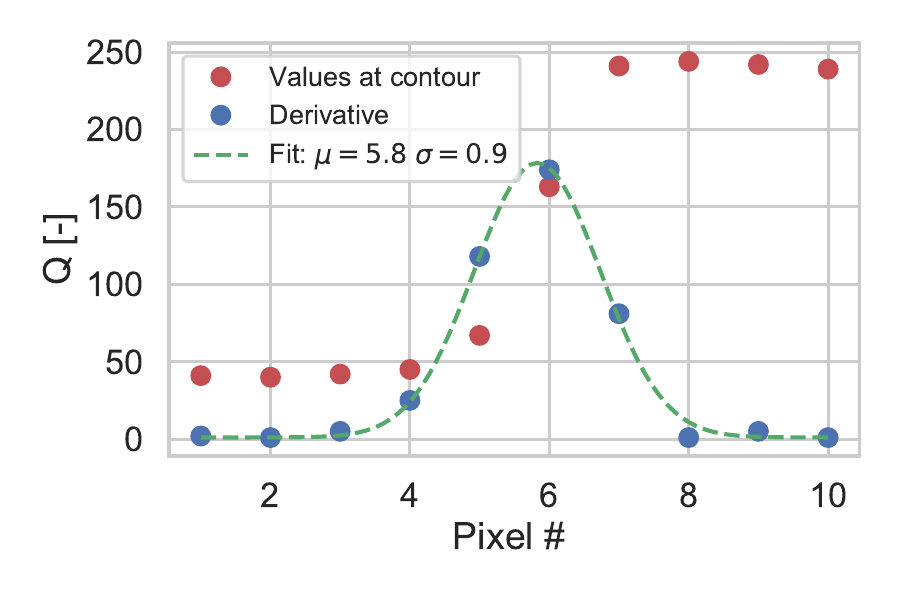}
		\caption{}\label{fig:contrast}
	\end{subfigure}
	\caption{(\subref{fig:pixcon}) A zoomed-in image of a vertical contour. (\subref{fig:contrast}) Red: the light
	intensity  $Q_{n}$ on a row of $n = 0-9$ pixels between values 0 - 255. Blue: differentiation: $Q_{n} = Q_{n+1} - Q_{n-1}$.}\label{fig:BlackWhiteCrossing}
\end{figure}

\textbf{The pattern of the coded mask} 

The position of the image on the sensor is defined by black-white contours of the pattern: see fig.~\ref{fig:pixcon}. The black-to-white transition is confined, in this case, by a row of $10$ pixels. The content of the pixels is displayed in the right-hand picture in red dots, forming a Fresnel edge diffraction curve. After differentiation, the top of the Gaussian-like curve through the blue dots is a measure for the contour position. In standard Rasnik systems (pixel pitch of $\SI{10}{\mu m}$, and sensor dimension of $4$ mm $\times$ $3$ mm), this contour position is determined with a precision of \textasciitilde$~\SI{0.2}{\mu m}$. This measurement is performed $15$k times along the total contour length of $\SI{100}{mm}$, both in the $x$ and $y$ directions, resulting in a spatial resolution of \textasciitilde $\SI{1.5}{nm}$ \textit{per image}.

Of all possible patterns, the chessboard pattern has the highest possible contour length in the $x$ and $y$ direction. Note also that the contour length is equal in both directions. The smaller the chess fields, the longer the total contour lengths. Because of diffraction, the ratio of the white and black pixel contents decreases with decreasing chess field size, approaching unity for small chess fields (gray image). This implies that there is an optimal chess field size.

The \textbf{range of measurement} of a Rasnik system is determined by the size of the coded mask: if the optical axis crosses a plain chess field coded mask, an analysable image will be generated. For a pure chess field pattern, the $x$ and $y$ response of the image analysis routine versus a mask displacement would be periodical. To break the periodicity, each ninth column and row of chess fields is coded digitally by inverting black and white, setting a valid bit. In this way, 8 bits are available for identifying the $9^{\text{th}}$, $18^{\text{th}}$, $27^{\text{th}}$, etc. row or column, and an arbitrarily large mask (and therefore dynamic range) can be realised.

\section{Error sources}\label{ErrorSources}
In this chapter the error sources limiting the performance of Rasnik are described in a comprehensive and complete way.
The error contributions are analysed or estimated. The quantum nature of light, causing fluctuations in pixel charge signals, is found to be the fundamental limit for spatial resolution. This is confirmed by the results given in section~\ref{Results}.

\subsection{The propagation of light through air}
Light rays are bent if the index of refraction $n$ of the ambient medium has a gradient in a direction perpendicular to the direction of the ray. This causes a first order error in the measured image position. In general this error is proportional to the square of the distance between the sensor and the mask.
The reduced index of refraction $(n - 1)$ of air is approximately proportional to the air specific density and thus pressure. Temperature gradients, and density fluctuations due to convection cause systematic shifts and variations with time in the measured image position. In addition, these variations may cause image deformation. The effects of density variations in ambient air are analysed in detail in~\cite{beker_2019}.

Temperature gradients in air can be reduced by shielding the light path with, for instance, metal tubes. Another way is to mix the ambient air by means of a set of fans. This causes, however, variations in air density, reducing the systematic error at the cost of an increase of the noise of the position measurement.

Another solution is to apply a medium with less variation in $n$. This could take the form of a glass or quartz bar with optical flat surfaces at the ends. These bars could be replaced by vacuum tubes equipped with optically flat windows at the ends. Finally, the complete system can be placed in vacuum, provided that heat dissipation, in particular of the image pixel sensor, is addressed. The vacuum should be such that the mean free path of the air molecules is in the order of the light beam diameter: a pressure of $\SI{1e-2}{mbar}$ is usually adequate.

\subsection{Errors in the coded mask}
The image analysis routine yields the $x$ and $y$ coordinate of the image on the sensor. Each of these coordinates is the mean of \textasciitilde $1.5 \times 10^4$ measured values of equal weight. As a result, a deviation of the nominal position of a section of a contour
contributes by a fractional weight to the response. In modern mask technology, contour positions can be made with a precision better than $\SI{6}{nm}$. With $1.5 \times 10^4$ samples, the contribution of the mask error to the statistical error in a measured coordinate is $\SI{50}{pm}$, assuming a Gaussian distribution of the contour position deviations.

Over a large range, the linearity of the $x$ and $y$ response is determined by the mask precision of the periodic pattern. The overall scale of a mask may depend on temperature, but by printing the mask on quartz or Zerodur~\cite{Veins_1990}, the scale error can be limited, if necessary. As a result, the linearity error of Rasnik can be as low as 50 pm over an arbitrarily large range provided that linearity errors within one ChessField period are dealt with.

\subsection{Image distortion}\label{distortion}

Optical image distortions cause perfect image points ($x$,$y$) to appear at ($x$',$y$'). The deviation is usually rotational symmetric around the optical axis.
An example is the well known pin-cushion effect. The deviations are constant, only depending on $x$ and $y$, and therefore do not cause a first order error since the $x$ and $y$ response is the average of $1.5 \times 10^4$ ($x$,$y$)-points of equal weight. These points are quite evenly distributed over the image area, so identical deviations are constantly present in each image. A third order error in linearity may occur in the case an image edge nearly coincides with a contour: 
in this case some parts of the contour participate, whereas others do not. This linearity error is periodic with the chess field size and with twice the chess field size. The error can be quantified by making density plots of real data with random $x$ and $y$, and an empirical correction can be applied.

\subsection{Image blurring}
Image blurring makes contours vague. The three contributions to image blurring are

\begin{figure}[htbp]
	\centering
	\begin{subfigure}[b]{0.49\textwidth}
		\includegraphics[width=\textwidth]{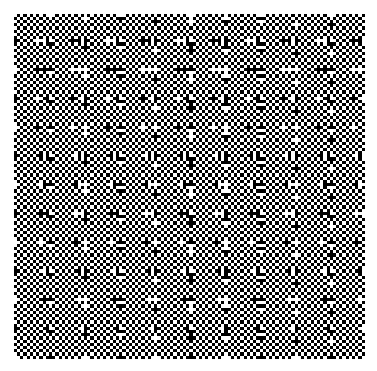}
		\caption{}\label{fig:ChessField}
	\end{subfigure}
	\begin{subfigure}[b]{0.49\textwidth}
		\includegraphics[width=\textwidth]{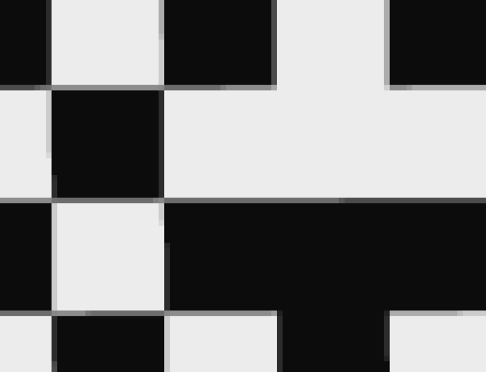}
		\vspace*{0.01cm}
		\caption{}\label{fig:pixbound}
	\end{subfigure}
	\caption{(\subref{fig:ChessField}) The ChessField pattern. Each 9th row and column includes its ID, providing the coarse position. (\subref{fig:pixbound}) Gray pixels: contour pixel crossings.}\label{fig:ChessFieldIntro}
\end{figure}

\begin{itemize}\label{blurr}
	\item the finite pixel size of the sensor. If a contour crosses a pixel its light content will be between high and low. As a result there will be a gray pixel sandwiched between light and dark pixels: see fig.~\ref{fig:pixbound}.
	\item diffraction. This fundamental optical effect is essential for the Rasnik performance. The black-white transition can be described by the Fresnel edge diffraction curve (see fig.~\ref{fig:contrast}). The steepness of the slope is inversely proportional to the wavelength $\lambda$ of the applied (led) light. The curve is non-symmetric, and a black-white transition is therefore not identical to a white-black transition.
	\item out-of-focus of sensor. This causes a linear transition from white to black. In practice,
	systems are well-focused. The effect could be used as fifth output parameter.
\end{itemize}

\subsection{Errors associated with the image pixel sensor}

\begin{figure}[htb]
\centering
\includegraphics[width=14cm]{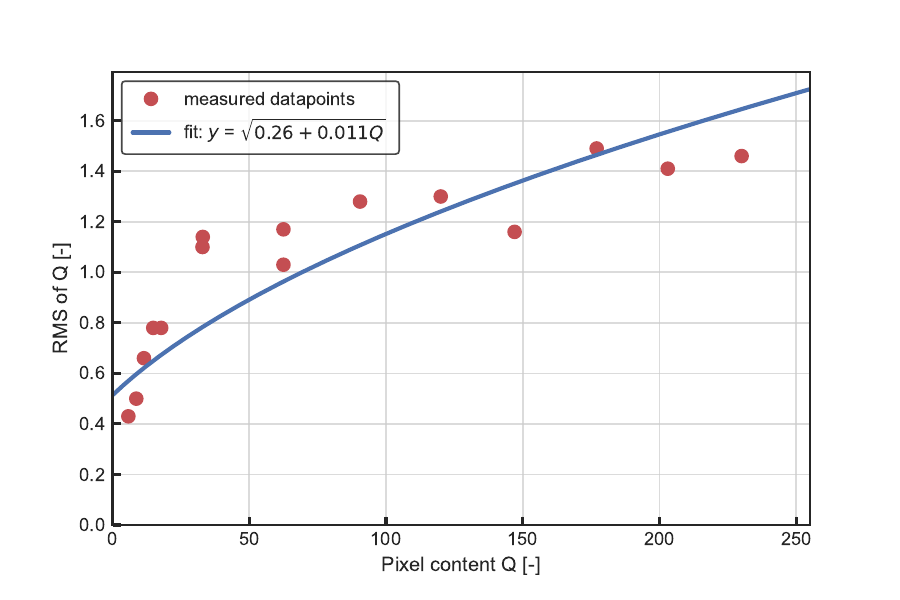}
\caption{The measured fluctuations in pixel content versus their average content of some selected pixels with different gray value.
For this, the 8-bit setting of the MER-041-302GM image sensor was used, and a set of 256 images was taken. The fluctuations of the pixel content has an electronic component and a component due to quantum fluctuations. The fit parameters are input for the image generator discussed in section~\ref{analysis}.}
\label{fig:quantfluc}
\end{figure}

\textbf{Quantum fluctuations: shot noise}. When light falls on the active area of a pixel of the sensor during the shutter open time, the absorbed photons create photo-electrons, charging up the well of the pixel: this charge is proportional to the light fallen onto the pixel. If $N$ photons are absorbed on average, the pixel electron content will fluctuate according a normal distribution with $\text{RMS} = \sqrt{N} $ due to the quantum character of light. 
In fig.~\ref{fig:quantfluc} the measured fluctuations are depicted for some selected pixels.
The amplitude of the fluctuations can be described as the sum of a squared constant term due to the
electronic (dark current) noise,
and a squared term proportional to the square root of the average pixel content, due to the quantum fluctuations.

The contribution of electronic noise is constant and equal for all pixels and is small with respect to the quantum fluctuations, as can be expected for modern pixel sensors. In the following, pixel noise includes electronic shot noise and quantum fluctuations.

In practice, the applied MER-041-302GM pixel sensor can operate with $8$-bit and $12$-bit readout. For light levels above 20 $\%$ of the maximum, the pixel noise is larger than the smallest 8-bit ADC step in pixel content. The 12-bit conversion, which reduces the maximum frame rate by a factor of order 2, does not improve the performance of the pixel sensor in our application.

The S/N ratio of the pixel content is improved when more light is absorbed. There is, however, an upper limit to the light level that CCDs or CMOS image pixels can accept before charge leaks away from the wells.
The amount of light, collected during the shutter time, resulting in the maximum ADC output of $255$ and $4095$ for the 8-bit and 12-bit conversion, respectively, matches the well capacity.
This limitation of CCDs and CMOS pixel image sensors can be reduced by applying sensors with a higher frame rate, but the development a special ASIC pixel sensor, capable to handle much higher light levels, would be the preferred solution.

\textbf{Pixel, row and column geometry}. Although image pixel sensors are mechanically very precisely made, thanks to state-of-the-art lithography, some pixel rows or columns may be wider than nominal, for instance. Such a deviation results in an constant image distortion as addressed in section~\ref{distortion}: the effect can be discarded.

\textbf{Pixel response: slope and pedestal, pixel cross talk}. A too large pixel content, due to a deviating slope and/or pedestal, may virtually pull or push the position of the nearest local contour, in the $x$ and $y$ direction. Since the direction of this virtual force has equal chance to be $-x$ or $+x$, respectively $-y$ or $+y$, the average force will be zero when taken the average of the deviations of $5 \times 10^5$ pixels. Pixel cross talk can be treated as a fourth source of image blurring, and is negligible with respect to diffraction.

\textbf{Pixel Flicker noise}. Since the image pixel sensor is a crystalline silicon solid, it may be considered as constant in time in terms of shape and dimension. This is also the case for the coded mask and the lens. A pixel output can be characterised in terms of a slope and pedestal. Flicker noise is due to changes in the depletion layer of pixels causing changes in characterisation and is one of the components of so-called 1/f, or pink, noise. This variation has little effect on the measured image position, as has been argued in the previous item. Even in the case of a common local group behaviour of pixels, the pulling or pushing of contours is randomised. Since $5 \times 10^5$ pixels participate, it will be a challenge to measure the practical Flicker noise, cleaned from temperature effects. Equivalently, the drift in Rasnik systems with time should be close to zero.

\subsection{Temperature effects}

\textbf{Mask.} If the temperature of the mask changes, for instance due to fractional absorption of the light emitted from the back-illuminating led, only the scale of the mask may change. The mask material should be selected in order to meet specifications. Masks can be printed on Zerodur with its thermal expansion coefficient lower than $10^{-9}$K$^{-1}$~\cite{Veins_1990}.

\textbf{Lens, or microscope objective.} Due to its transparency, the temperature of a lens will change little. Moreover, this does not cause a variation in the position of the optical centre. There may be a small variation of the focal length, but this does not change the image scale. The image blurring may vary slightly, but this is a negligible third order effect, leaving the output data unaffected.

\textbf{Image pixel sensor.} The silicon chip will expand by $\SI{2.6e-6}{K^{-1}}$ with temperature. State-of-the-art CCDs or CMOS sensor chips dissipate little, while the signal processing circuitry may be positioned elsewhere, including cooling. Thermal expansion causes an error in the measured scale of an image, and can be minimised by gluing the sensor chip, in its centre, at the flat tip of a cooled Zerodur rod. The position of the optical centre of the sensor remains unaffected, and a change of temperature of the sensor has no influence on the x and y data.

\textbf{Component suspension, arm effects.} The position information of mask, lens and sensor is to be transferred to the objects whose alignment is to be monitored. There may be three component holders that may be subject to thermal expansion, and variations appear as drift in the measured $x$ and $y$ coordinates. These variations can be minimised by integrating the Rasnik components carefully in the design of the application.

\begin{figure}[htbp]
	\centering
	\begin{subfigure}[b]{0.49\textwidth}
		\includegraphics[width=\textwidth]{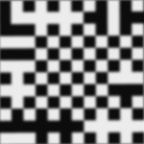}
		\caption{}\label{fig:blurr1}
	\end{subfigure}
	\begin{subfigure}[b]{0.497\textwidth}
		\includegraphics[width=\textwidth]{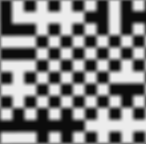}
		\caption{}\label{fig:blurr2}
	\end{subfigure}
	\caption{(\subref{fig:blurr1}) Image blurring: 2D Gauss width $\sigma = \SI{6.0}{\mu m}$.
	(\subref{fig:blurr2}) $\sigma = \SI{12.0}{\mu m}$}\label{fig:label5}
\end{figure}

\begin{figure}[htb]
\centering
\includegraphics[width=8cm]{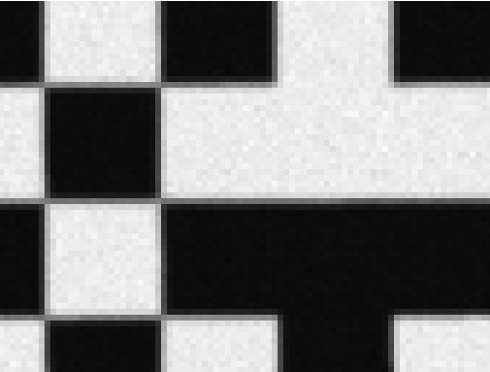}
\caption{Exaggerated pixel content noise}
\label{fig:ImaNoise}
\end{figure}

~\newline        
\noindent\fbox{%
    \parbox{0.975\textwidth}{%
        \textbf{The spatial resolution of Rasnik systems is limited by quantum fluctuations in the applied light.}
    }%
}
~\newline   

\section{Image simulations and image analysis}\label{analysis}

\begin{table}[htb]\caption{The parameters applied in section \ref{analysis}}.
\centering
\begin{tabular}{| l | l |l |}
\hline
Parameter & unit & description \\
\hline
$x$         & $\mu$m &$x$-coordinate of crossing point of mask and optical axis also output\\
 & & \hspace{0.5cm}of the SOAP routine\\
$y$         & $\mu$m & $y$-coordinate as x above  \\
$x_{\text{sim}}$ & $\mu$m & $x$-coordinate input for the image generator  \\
$y_{\text{sim}}$ & $\mu$m & $y$-coordinate input for the image generator  \\
$\theta_z$      & rad & Differential rotation of mask and pixel sensor. For $\theta_z$ = 0, the\\
 & & \hspace{0.5cm}grids of sensor and mask are in line\\
S         & (-) & Image scale, equal to ratio of image distance and object distance  \\
Br        & $\mu$m &Image blurring constant: sigma (radial) of $2$D Gaussian distribution   \\
 & & \hspace{0.5cm}over the pixel area\\
\hline
\end{tabular}
\label{parameters}
\end{table}

A Rasnik system produces image data files which are sequentially processed by an image analysis program. The output of the analysis program are floats representing $x$, $y$ and $\theta_z$.
A simulation program was developed which generates image data files. The only variation in the image data lies in the pixel content: the fluctuations, set out in fig~\ref{fig:quantfluc}, were generated by means of a MonteCarlo (MC) method.
The MC generated images were processed by the image analysis program, and the output parameters were compared to the input parameters of the MC simulation program. In the case of agreement, the performance of both the MC program and the analysis program can be obtained. In the case of large deviations it may not be clear whether these origin in the MC program or in the analysis program.

The parameters used throughout this section are listed in Table \ref{parameters}.

\textbf{Image simulations.}
First, the typical ChessField pattern is generated (fig~\ref{fig:ChessField}). For a given $(x,y)$ as input, a $0$ (black) or $1$ (white) is returned. In a next step the pixel response is calculated: by sampling an array of points within the pixel area, the fractional pixel content is calculated, relevant in case a contour crosses a pixel area (fig~\ref{fig:pixbound}). If relevant, the inter-pixel dead area can be taken in account by reducing the fiducial pixel area. Then, image blurring is addressed by spreading the pixel content over an array of the neighbouring $11\times11$ pixels following a 2D Gaussian distribution; this area is large enough to cope with very wide distributions. The width of this distribution is obtained from real data. The simulated image blurring is shown in figs~\ref{fig:blurr1} and~\ref{fig:blurr2} for two different Gaussian widths. Finally, fluctuations in the pixel content are applied following the relation $Q_{\rm{fluc}} = \sqrt {A + B.Q}$ found in fig~\ref{fig:quantfluc}, with  $A = 0.26$ and $B = 0.011$.
Fig~\ref{fig:ImaNoise} shows the effect of exaggerated shot noise in the pixel contents.

\begin{figure}[htbp]
	\centering
	\begin{subfigure}[b]{0.565\textwidth}
		\includegraphics[width=\textwidth]{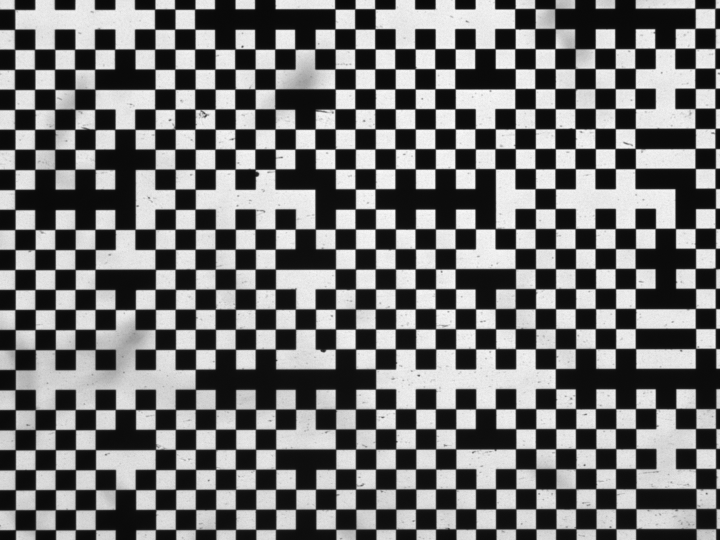}
		\caption{}\label{fig:ima}
	\end{subfigure}
	\begin{subfigure}[b]{0.28\textwidth}
		\includegraphics[width=\textwidth]{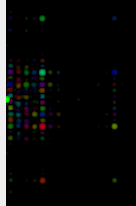}
		\caption{}\label{fig:fft}
	\end{subfigure}
	\caption{(\subref{fig:blurr1}) Typical Rasnik image from SolidRas. Here, a ChessField mask of $\SI{120}{\mu m}$ was applied.
	(\subref{fig:blurr2}) Typical SolidRas FFT-converted Rasnik image of fig.~\ref{fig:ima}. The peaks are associated with the periodicity of the image pattern defined by the ChessField side and by the coarse rows and columns at 9 times the ChessField side. All peaks may contribute to the determination of the phase of the contours.}
\end{figure}

\textbf{Image analysis.}
The SOAP image analysis program for processing the ChessField Rasnik images is set out in detail in \cite{beker_2019}.
First, the raw image as depicted in fig~\ref{fig:ima} is converted into a 2D Fourier image, shown in fig~\ref{fig:fft}.
Here, the phase of each Fourier peak is coded by colour.
The positions of the strongest (primary) peaks reveal the frequency and
possible rotation of the pattern, while the complex phases indicate a shift.
Secondary (harmonic) peaks can be used to refine these parameters, and
the nominal positions of the squares is found. Then the
intensities near the centers of the squares is evaluated to determine
the positions of the code lines, and the values encoded in them.

\textbf{Verification of the SOAP image analysis program.} The performance of the SOAP analysis program has been certified to be better than $\SI{1}{nm}$ in resolution and $\SI{10}{nm}$ in linearity~\cite{beker_2019}. With the image simulation program, pgm-formatted image files are generated: after analysing these by SOAP, the output can be compared to the input parameters of the simulation, revealing systematic and statistical deviations. If significant deviations are found, however, it may not be clear if these originate from the simulation or from the image analysis routine.

In the following, the image data generator uses parameters (i.e. blurring and geometry) of the SolidRas system,
described in detail in section~\ref{Results}.

\begin{figure}[htb]
\centering
\includegraphics[width=13cm,height=9cm]{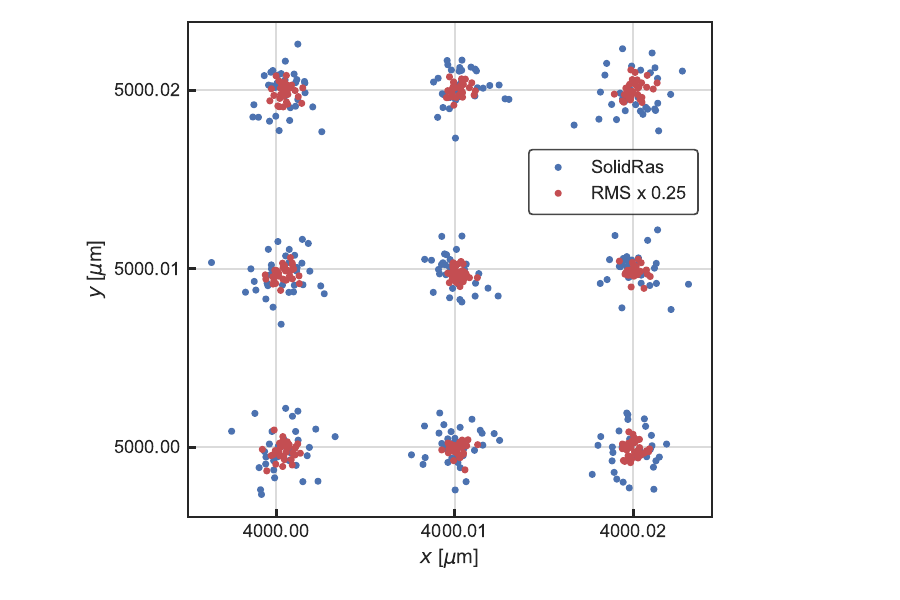}
\caption{The $(x,y)$ response of the SOAP image analysis routine for images from a SolidRas MonteCarlo generator.
The $(x,y)$ input of the generator was $(4000, 5000)$ for the bottom-left point,
and was followed by 8 points with steps of $\SI{0.01}{\mu m}$ and $\SI{0.02}{\mu m}$, in $x$ and $y$, respectively.
The non-linearity is noticeable; most likely this is caused by the phase determination of the Fourier peaks.
The spread associated with a point is only due to fluctuations in the pixel content.
Blue: normal SolidRas parameters; red: pixel quantum fluctuations reduced by factor $0.25$.}
\label{fig:xy}
\end{figure}

\textbf{Spatial resolution.} The quality of the SOAP image analysis program, in terms of spatial resolution and linearity, has been verified by the production of a series of SolidRas images, generated at nine $(x_{\text{sim}} , y_{\text{sim}})$ positions, with the first (bottom-left) position at $(\SI{4000}{\mu m} , \SI{5000}{\mu m})$.
The other $(x_{\text{sim}},y_{\text{sim}})$ positions lie on a square $10$ by $\SI{10}{nm}$ grid.
The $(x,y)$ output of the SOAP analysis program is shown in fig.~\ref{fig:xy} by the blue points.
The spatial resolution is of order $\SI{1.5}{nm}$, equal for the $x$ and y. The red dots show the effect of reducing the amplitude of the quantum fluctuations by a factor $0.25$. The output is scattered in a continuous fashion, instead of yielding only some discrete levels: the image analysis routine appears to allow sub-nm level in spatial resolution.

\begin{figure}[htb]
\centering
\includegraphics[width=14cm]{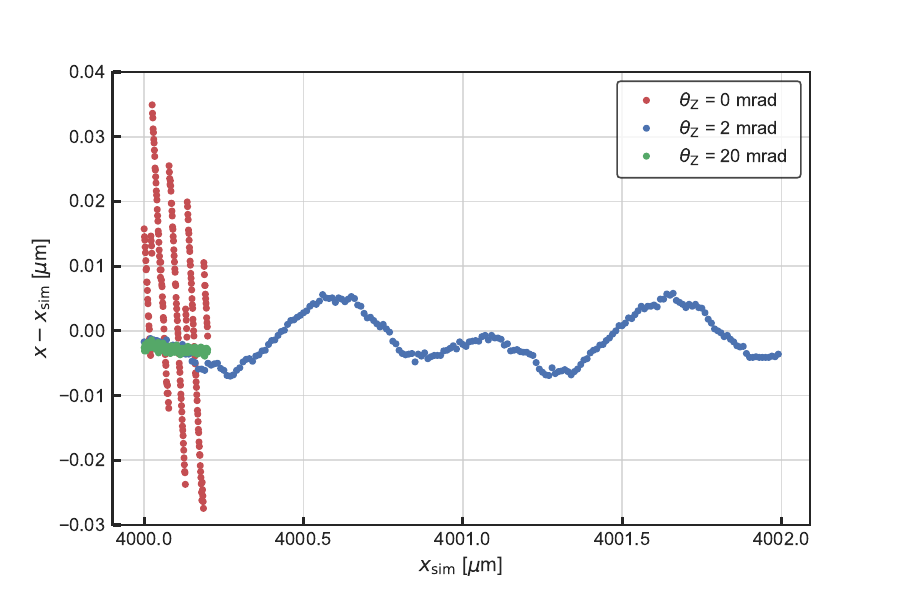}
\caption{Linearity error of the combined MonteCarlo image generator and the SOAP image analysis program. The amplitude of the error depends strongly on the differential rotation $\theta_z$ of the coded mask and image pixel sensor around the $z$-axis.}
\label{fig:linea}
\end{figure}

\textbf{The linearity of the response.} The systematic deviation of the SOAP analysis output ($x,y$) from the associated generator input values $(x_{\text{sim}} , y_{\text{sim}})$ is clearly visible in fig.~\ref{fig:xy}. The output for $(x_{\text{sim}},y_{\text{sim}})=(\SI{4000}{\mu m},\SI{5000}{\mu m})$ is centred around $(x,y)=(\SI{4000.001}{\mu m},\SI{4999.999}{\mu m})$ instead of being centred around the input.

The linearity of the SOAP analysis routine is tested by generating images with a step-wise variation in $x_{sim}$, while keeping $y_{sim}$ fixed, fluctuations in pixel content small, and $\theta_z$ set at zero.
In fig.~\ref{fig:linea} the difference between the generator input $x_{\text{sim}}$ and the average analysis output $x$ is plotted against the input $x_{\text{sim}}$.
The result of the analysis program deviates up to $\SI{35}{nm}$ over a range in $x_{\text{sim}}$ of $\SI{50}{nm}$.
This would result in a deviation of the local slope (d$x$/d$x_{\text{sim}}$) of $70$ percent or worse.
In green, the deviation is shown with $\theta_z=\SI{20}{mrad}$: the linearity error is greatly reduced, and evolved in an offset error of order $\SI{3}{nm}$. This suggests that the error is associated with the determination of the (Fourier output) phase. This deviation needs to be studied, and possibly other algorithms than Fourier conversion must be developed. In blue dots, the linearity error for a range of $\SI{2}{\mu m}$ and $\theta_z=\SI{2}{mrad}$ is plotted, indicating that the magnitude of the error is limited to $\SI{5}{nm}$ in practical cases.

\textbf{Data processing.} As readout system for the MER-$041$-$302$ (Ethernet) image pixel sensor, a Dell OptiPlex $3020$ was used with Linux Ubuntu $18.04.4$ LTS as operating system. A GPU (GeForce GTX $1050$/PCIe/SSE$2$) card was added to speed up Fourier conversion. With this, a frame rate of $\SI{275}{Hz}$ was possible, albeit that four pixels in a square were were combined into one superpixel. This reduced the time of the raw image data transfer by a factor $4$.

\section{Results of the SolidRas and MicroRas systems}\label{Results}
\begin{figure}[htb]
\centering
\includegraphics[width=12cm]{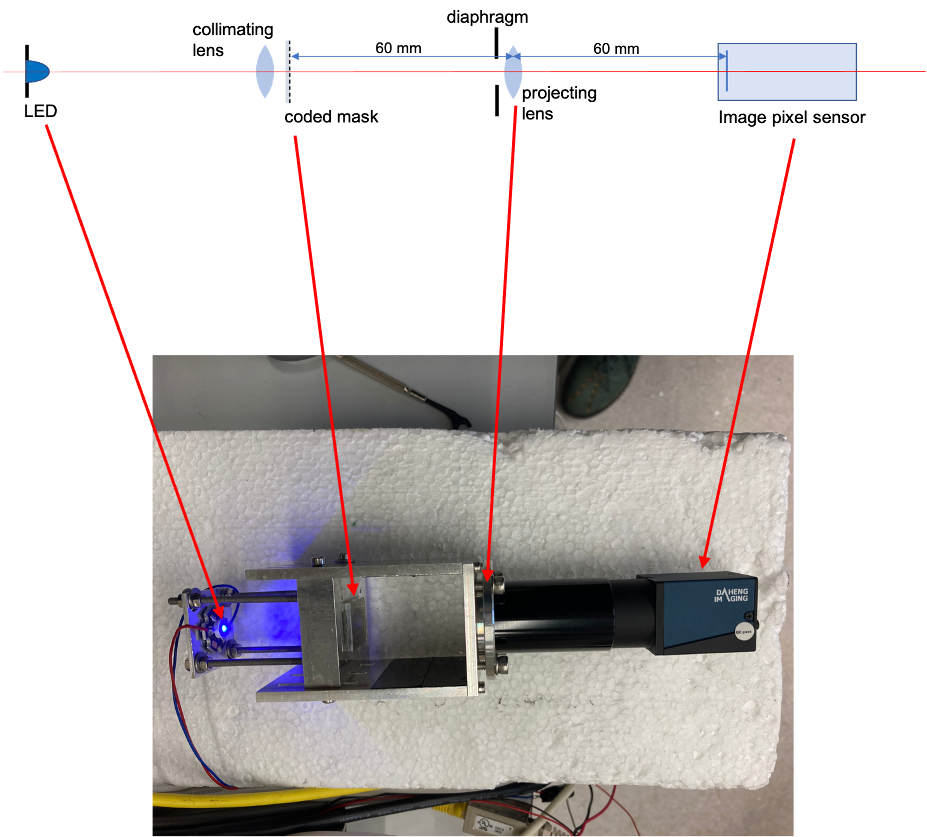}
\caption{Layout of the SolidRas system: although quite solid and stiff, the flexibility in the plane of the picture is less than in the normal of the plane due to the U-profile on which the lens and light source are fixed. The eigenfrequencies, appearing in fig.~\ref{fig:solidbest}, in the $x$ direction are therefore much higher.}
\label{fig:solidraslayout}
\end{figure}

SolidRas is an assembly of light source, coded mask, bi-convex lens and image pixel sensor, fixed together in a stable and solid unit, see fig.~\ref{fig:solidraslayout}. It provides image data with little variation, and the unit can be suspended as pendulum, reducing the effect of external (seismic, acoustical) vibrations.

The LED (Luxeon StarLed RoyalBlue LXHL-MRRA,  $\lambda = \SI{448}{nm}$), and collimating lens (Edmund Optics \#$63-539$) form a K\"ohler light source, providing efficient back-illumination of the coded mask~\cite{kohler}. The ChessField mask has chess field sides of $\SI{85}{\mu m}$. The projecting lens, identical to the collimating lens, has a focal length of $\SI{30}{mm}$. During the data acquisition presented below, a diaphragm of $\SI{8}{mm}$ diameter was applied. As image pixel sensor, the Daheng Mercury MER-$041$-$302$G was applied with the following specifications: $720\times540$ pixels, pixel pitch sq. 6.9 ${\mu}$m, 8-bit deep conversion, max frame rate 302 fps.

\begin{figure}[htb]
\centering
\includegraphics[width=12cm]{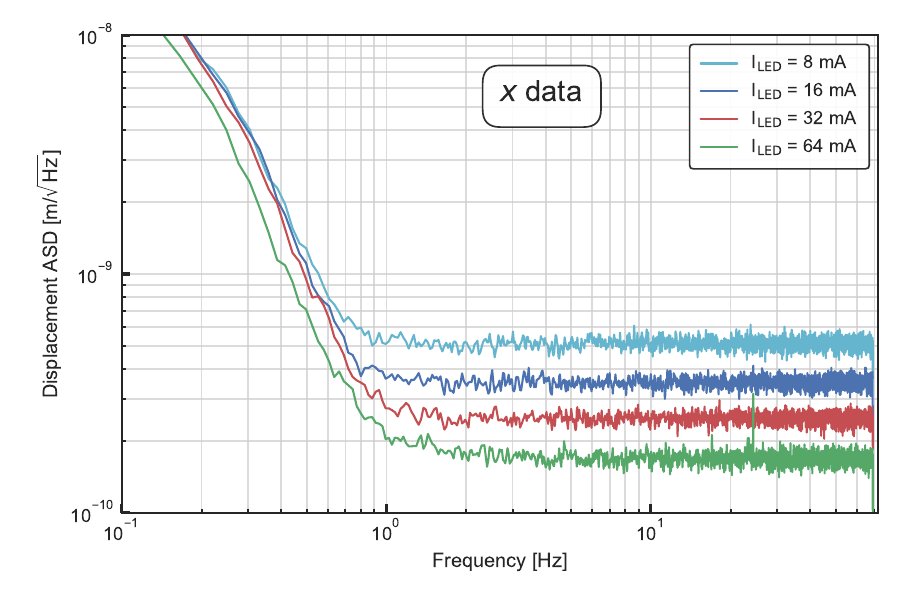}
\caption{The noise floor levels for maximum light intensity, and for the LED current reduced by a factor 2, 4 and 8.}
\label{fig:halfcurrents}
\end{figure}

\begin{figure}[htbp]
	\centering
	\begin{subfigure}[b]{1.0\textwidth}
		\includegraphics[width=\textwidth]{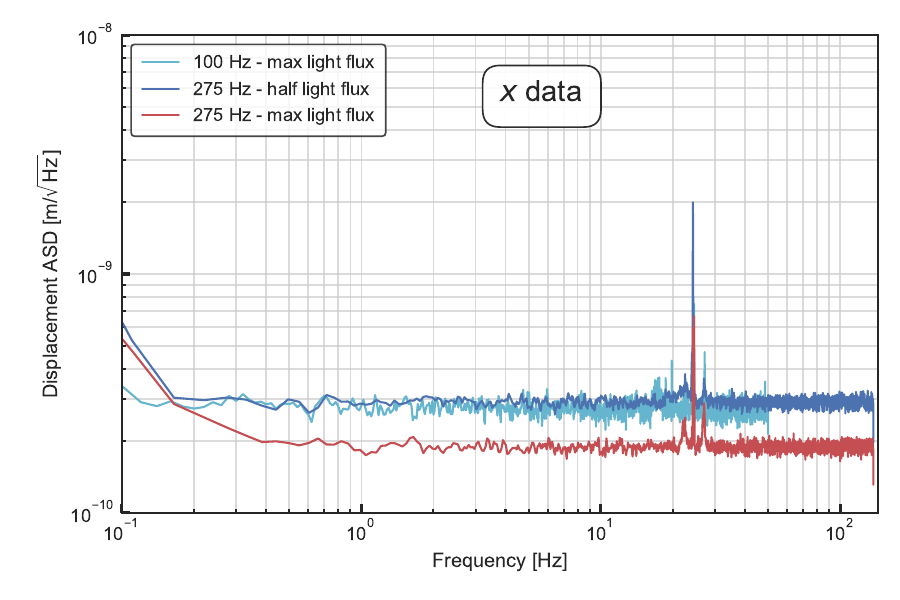}
		\caption{}\label{fig:3plotsx}
	\end{subfigure}
	\begin{subfigure}[b]{1.0\textwidth}
		\includegraphics[width=\textwidth]{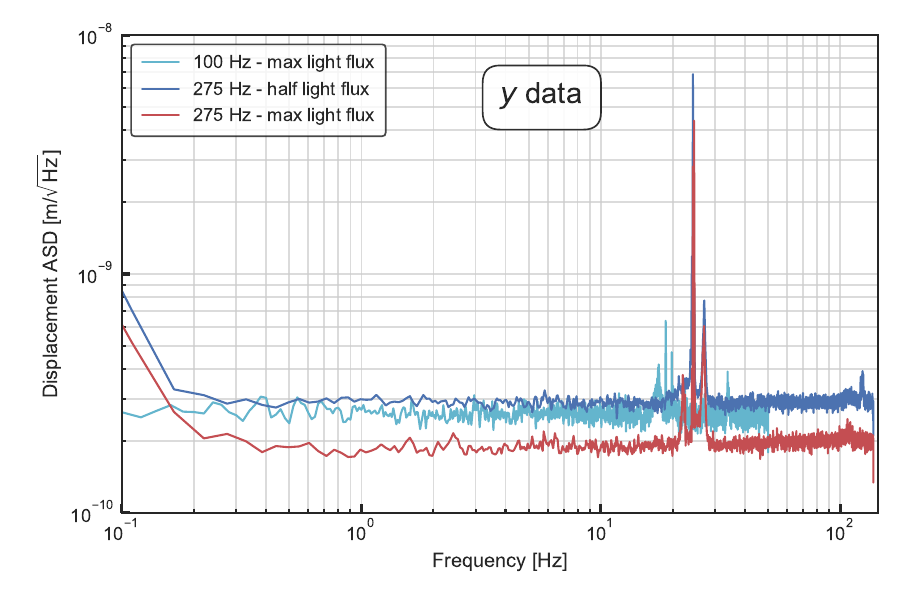}
		\caption{}\label{fig:3plotsy}
	\end{subfigure}
	\caption{Welch plot of a SolidRas data set, taken during a period of 30 minutes. Red: best performance at maximum practical frame rate (275 Hz). Light blue: frame rate reduced to 100 Hz, maintaining exposure time. Dark blue: frame rate 275 Hz, LED current reduced by factor 2. (\subref{fig:3plotsx}) $x$ data, and (\subref{fig:contrast}) $y$ data.}
\label{fig:3plots}
\end{figure}

 The SolidRas unit can be placed in vacuum and suspended as pendulum by two thin wires in order to limit mechanical variations induced by seismic and acoustic vibrations. Data is taken using an internal trigger setting the frame rate. The real-time image analysis yields a data file which includes a time stamp, the $x$ coordinate, the $y$ coordinate,
 the image scale (\textasciitilde 1), and $\theta_z$.

\textbf{Welch plots.} In order to obtain the contribution of various noise sources, a Power Spectral Density (PSD) plot or Amplitude Spectral Density (ASD) plot is desired.
For these parameters a Fast Fourier Transform can be executed in order to obtain the ASD.
In this paper the Welch's Method~\cite{welch} is used, in which the data is divided into overlapping windows, calculating the periodogram of each window, followed by taking their average. In this paper Hann's window, a variation of the rectangular window function, is used. The Hann window has an overlap $N_{\rm{overlap}}=N_{\rm{perwin}}/2$, corresponding to an overlap of 50\%. The square root is taken of the output of Welch's PSD estimation to obtain the ASD in $\SI{}{m/\sqrt{Hz}}$. The relevant parameters to be considered are: quantity of $x$ and $y$ data, sample frequency (= frame rate) and number of windows. The data sets spanned one hour, and were divided in 5000 windows.

The performance of SolidRas, in open air on a lab table, is shown in fig.~\ref{fig:halfcurrents}. The noise increase at low frequencies can be attributed to variations in the index of refraction of the ambient air. A noise floor of $\SI{180}{pm/\sqrt{Hz}}$ is reached when the
intensity of the LED light is adjusted such that the maximum pixel content does not exceed the 8-bit content (255), see the green curve. For the red curve, half the LED current was used, and for the light blue and dark blue curves, again half of the current was applied, respectively. The noise floor level rises a constant factor $\sqrt{2}$ when the light amplitude is reduced by a factor 2. This tallies with the consideration that the data of two images can be superimposed, forming an image with twice the light intensity. This demonstrates that the spatial resolution of Rasnik is limited by the quantum fluctuations in the pixel contents Q and that other noise sources such as pixel dark current, contribute little.

Next, the influence of the frame rate on the performance of the pendulum-suspended SolidRas is evaluated. The corresponding ASD is shown in fig.~\ref{fig:3plots}: the bottom (red) curve shows the best performance, taken with (practical maximal possible) $275$ frames per second (fps). The exposure time was set as high as possible. The current through the LED was adjusted such that the maximum pixel content was just below 255, which is the maximum digital output for the 8-bit setting. The electric power transferred to the led, for that setting, was $\SI{300}{mW}$. According the manufacturer, $20$ percent of this power is emitted through photons. Taking into account the optical transmission efficiency, the quantum efficiency of the pixel sensor and the shutter time, the number of photons onto a pixel agree with the well depth (22k $e^{-}$) of this pixel sensor.

The light blue curve shows the result of data taken at a frame rate of $\SI{100}{Hz}$ while keeping the exposure time unchanged. The spatial resolution \textit{per image} remains unchanged. The noise floor level is increased by a factor of $\sqrt{275/100}=1.6$ as expected.
The light output of the LED was verified to be well-proportional to its current. The dark blue curve, taken at the maximum frame rate of $\SI{275}{Hz}$ shows the effect of the current, thus light intensity, reduced by a factor $2$, as was observed in fig.~\ref{fig:halfcurrents}.
This demonstrates again that the fundamental limitation of Rasnik's spatial resolution is due to white noise in the pixel signals, and that the measured $x$ and $y$ have a Gaussian distribution.

The spatial resolution per image has been obtained by the analysis of Monte-Carlo generated images. The settings of the simulation program were adjusted such that the difference between simulated and real images was minimised: 

\begin{itemize}
	\item a pixel content dark region of $25$;
	\item a pixel content light region: $252$ (just below the $8$-bit maximum of $255$);
	\item a width of Gauss distribution describing image blurr: $\SI{4.5}{\mu m}$ (see fig.~\ref{fig:pixcon} and section ~\ref{blurr}). This value was obtained after analyzing the distributions of pixel row content covering contours (see fig~\ref{fig:BlackWhiteCrossing}), and is in good agreement with the expected blurring for a  diaphragm of $\SI{4}{mm}$ in this diffraction-limited case;
	\item The quantum fluctuations of pixel content $Q$ were taken as $\sigma_{Q} = \sqrt{0.26 + 0.011 Q}$, as found in fig.~\ref{fig:quantfluc}.
	
\end{itemize}
	
\begin{figure}[htb]
\centering
\includegraphics[width=10cm]{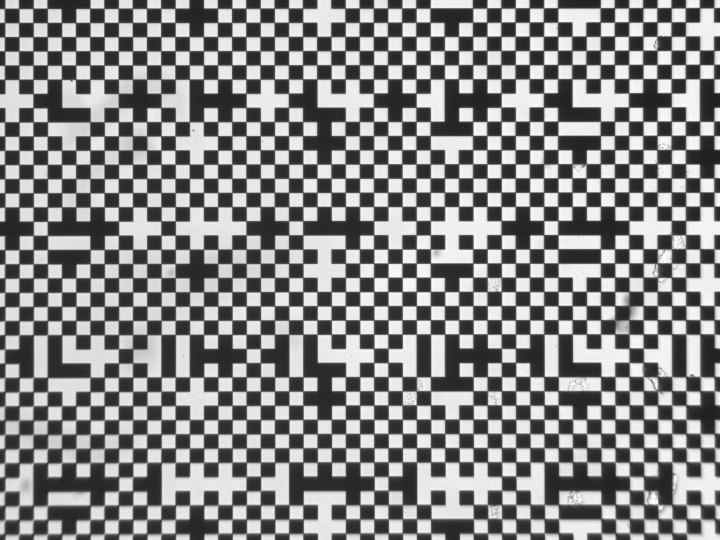}
\caption{Real SolidRas image, displaying some stains and dust particles.}
\label{fig:realsolidras}
\end{figure}

\begin{figure}[htb]
\centering
\includegraphics[width=10cm]{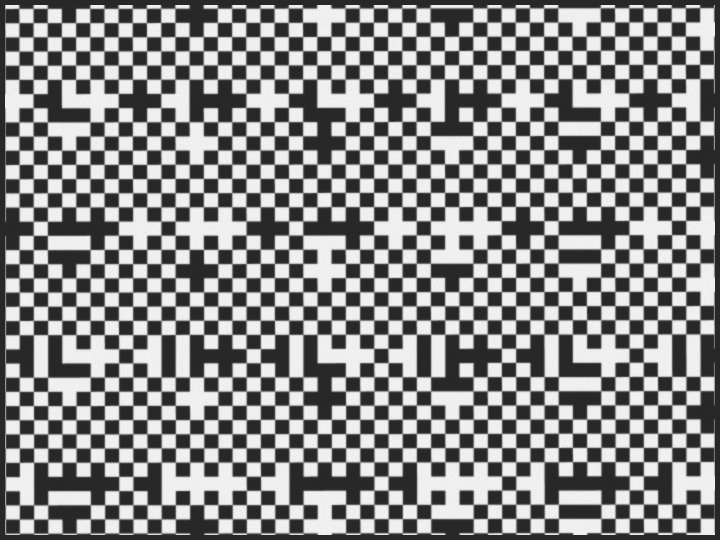}
\caption{The image of fig~\ref{fig:realsolidras} simulated by the Monte Carlo (MC) generator.}
\label{fig:MCSolid}
\end{figure}

A set of $256$ images was generated for which each pixel the quantum fluctuation was superimposed onto the pixel content.
A real and a simulated image is shown in figs.~\ref{fig:realsolidras} and~\ref{fig:MCSolid}. After analysis with SOAP, a spatial resolution per image  of $\SI{3.0}{nm}$, both in $x$ and $y$, was obtained. For the half-current data, this value was $\SI{4.8}{nm}$.

Frame rates above $\SI{100}{Hz}$ were only possible with reduced data formats: four pixels in a square were combined into one pixel, with $8$-bit depth. The simulations result in spatial resolutions per image of $\SI{3.3}{nm}$ for full LED current, and $\SI{5.0}{nm}$ for half the LED current, both in $x$ and $y$. The spatial resolution associated with superpixels is expected to worsen a little since a black-white transition is still covered by $3$ to $4$ pixels. The noise floor level following from the simulations should equal $\SI{3.3}{nm}/\sqrt{\SI{275}{Hz}}$ = $\SI{200}{pm/\sqrt{Hz}}$ in agreement with fig.~\ref{fig:3plots}.
This spatial resolution was predicted by Cramer-Rao simulations before \cite{JorisMSc}.

\begin{figure}[htbp]
	\centering
	\begin{subfigure}[b]{0.49\textwidth}
		\includegraphics[width=\textwidth]{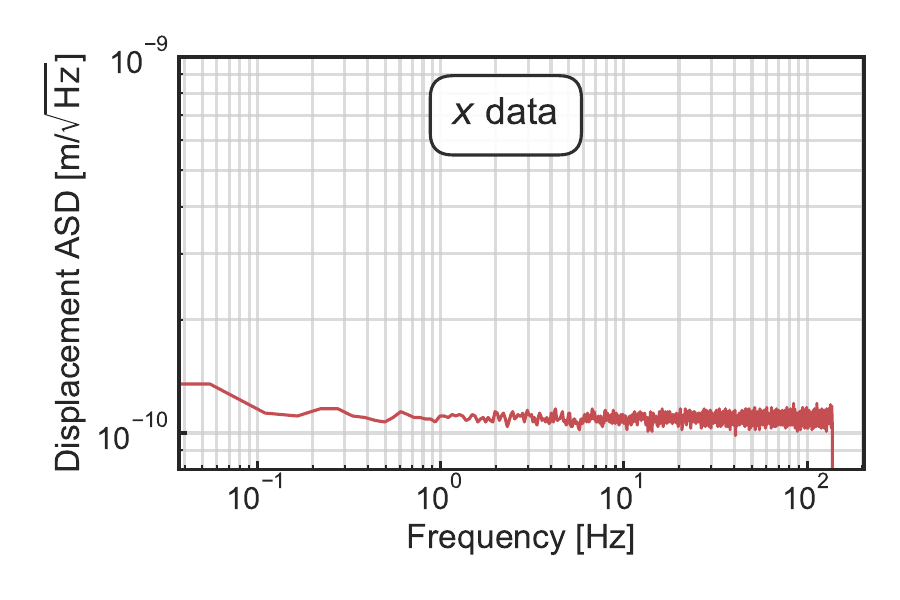}
		\caption{}\label{fig:solidx}
	\end{subfigure}
	\begin{subfigure}[b]{0.49\textwidth}
		\includegraphics[width=\textwidth]{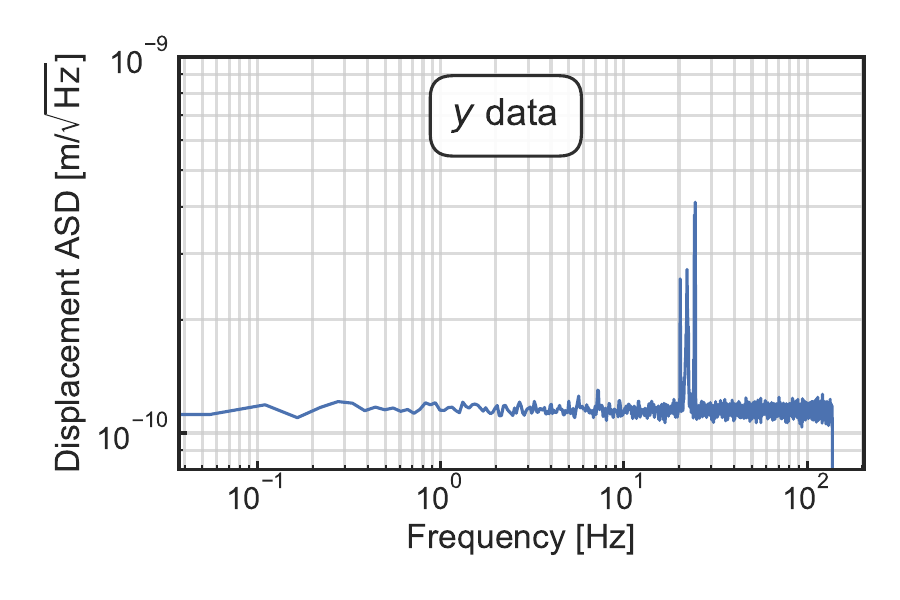}
		\caption{}\label{fig:solidy}
	\end{subfigure}
	\caption{Welch plots of continuous SolidRas $x$ and $y$ data. Diaphragm = $\SI{8}{mm}$, $\lambda = \SI{365}{nm}$.}
\label{fig:solidbest}
\end{figure}

In a next step, the LED was replaced by the UV LightAvenue KE$120$UYG, reducing the applied wavelength from $448$ to $\SI{365}{nm}$.
In addition, the diameter of the diaphragm was doubled from $4$ to $\SI{8}{mm}$. The image blurring, and, subsequently, the spatial resolution is expected to be improved by a factor 2, by just the latter change.
To test the performance of SolidRas unit with these changes, it was carefully suspended,
in a vacuum tank, by means of two thin polyester strings with a length of \textasciitilde$\SI{70}{mm}$.
The displacement-ASD result is shown in fig.~\ref{fig:solidbest} where a noise floor of $\SI{110}{ pm/\sqrt{Hz}}$ is reached. The peaks in the $y$ data in the $\SI{20}{Hz}$ region can be associated with longitudinal vibrations of the suspension strings. The difference in flexibility of the SolidRas unit in the $x$ and $y$ directions explains the absence of these peaks in the $x$ data: the associated resonance frequencies are much higher and damped more strongly.

The measured noise floor level for the Rasnik output parameter $\theta_z$ is
$\SI{1.0e-6}{rad/\sqrt{Hz}}$, in good agreement with simulations.

\begin{figure}[htb]
\centering
\includegraphics[width=12cm]{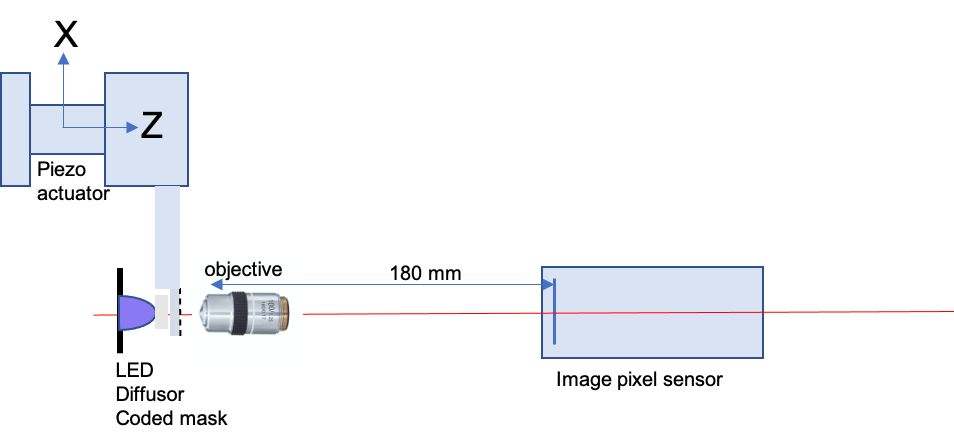}
\caption{The layout of the MicroRas system. The mask can be displaced in $x$,
$y$ and $z$ by manual dial gauges and, in addition, in $x$ and $z$  by means of piezo-actuators.}
\label{fig:microraslayout}
\end{figure}

\begin{figure}[htb]
\centering
\includegraphics[width=12cm]{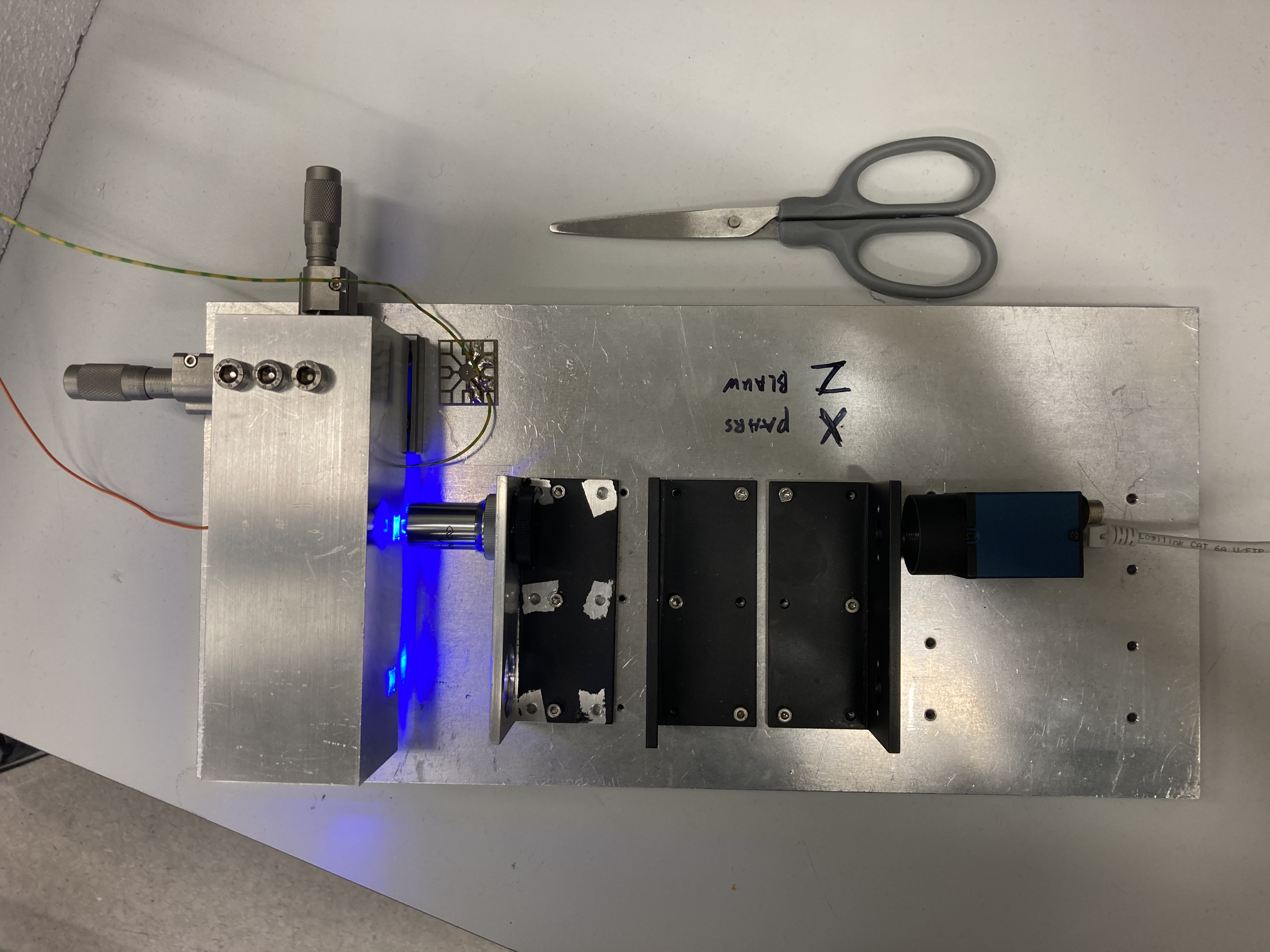}
\caption{The MicroRas unit: the MER pixel sensor, microscope objective and mask suspension are 
bolted firmly onto a common base plate.}
\label{fig:microras}
\end{figure}

\begin{figure}[htbp]
	\centering
	\begin{subfigure}[b]{1.0\textwidth}
		\includegraphics[width=\textwidth]{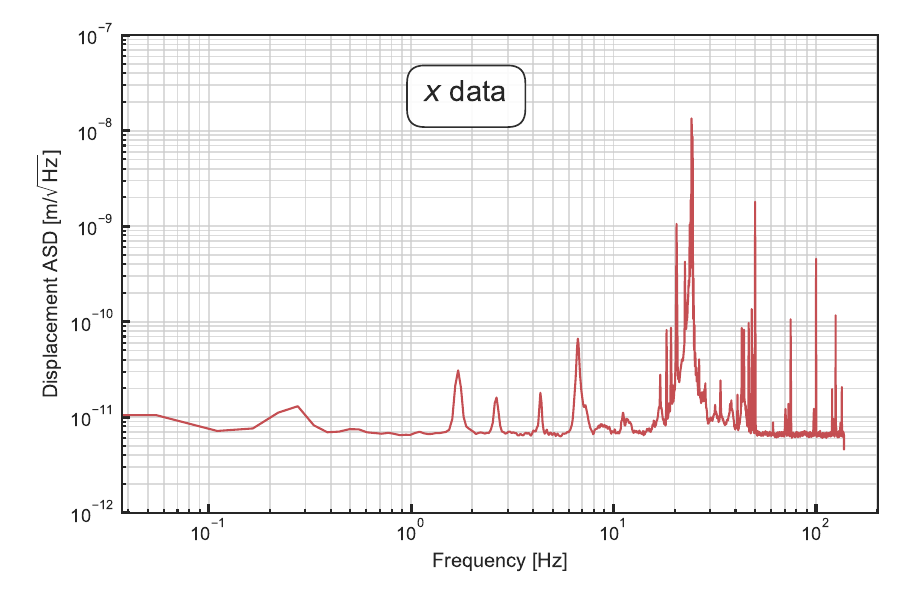}
		\caption{}\label{fig:microx}
	\end{subfigure}
	\begin{subfigure}[b]{1.0\textwidth}
		\includegraphics[width=\textwidth]{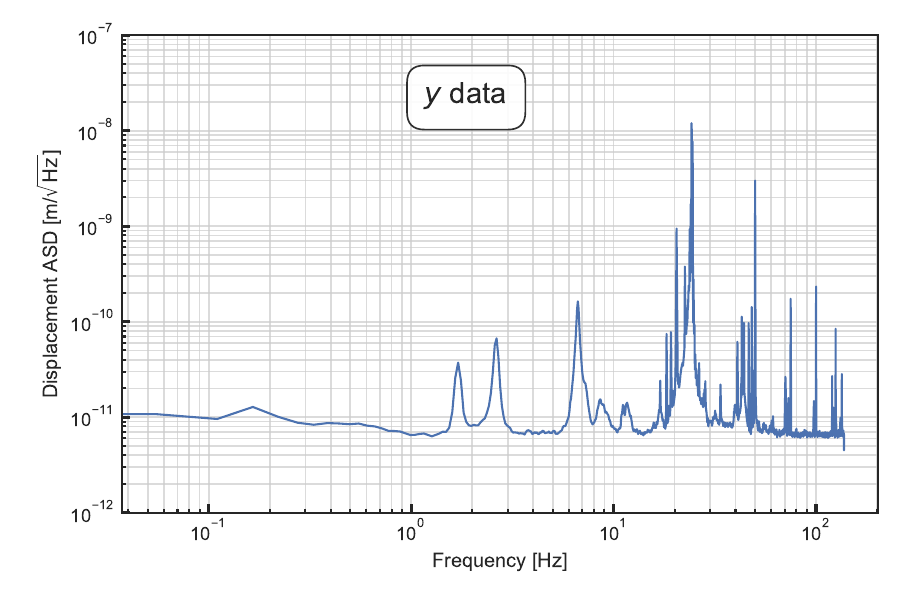}
	\end{subfigure}
    \caption{Welch plots of continuous $x$ and $y$ data of the MicroRas system.
    The frame rate is 275 Hz.}\label{fig:microras_welch}
\end{figure}

\textbf{Performance of the MicroRas system.} The layout of the MicroRas system is shown in figs.~\ref{fig:microraslayout} and ~\ref{fig:microras}. Instead of the projecting lens as in SolidRas, a microscope objective (Newport M-$20$X) is applied: the image displacement equals $20\times$ the mask displacement, improving the performance. The coded mask ($\SI{5}{mm}\times\SI{5}{mm}$)~\cite{photronics} is back-illuminated by means of a glued-on diffusor (Opaline $\SI{3}{mm}$)
and a LED (Luxeon StarLed RoyalBlue LXHL-MRRA $1$W) with $\lambda = \SI{448}{nm}$. The led-mask unit is mounted onto a $3$D stage in which $x$ and $y$ can be adjusted manually by means of dial gauges.
In addition, the $x$ and $z$ position can be remotely adjusted by means of a piezo-actuator (PiezoSystemJena HPS$1000$/$25$-$15$/$7$): this enables the necessary adjustment of the optimal in-focus position of the mask after applying vacuum, to a maximum of $\SI{9}{\mu m}$. The MicroRas system is suspended by two thin $\SI{30}{mm}$-long polyester wires attached to the top left and right corners of the base plate.

The performance of the MicroRas system is shown in fig.~\ref{fig:microras_welch}. Peaks near $\SI{3}{Hz}$ may be associated with the suspension mechanics. The wide peak at $\SI{25}{Hz}$ may be associated with eigenfrequencies of the unit itself. The sharp peaks at $25$, $50$, $75$ and $\SI{100}{Hz}$ may be associated with grid power electro-magnetic signal interference, or with grid power induced Eddy currents deforming the MicroRas base plate. A noise floor level of $\SI{7}{pm/\sqrt{Hz}}$ is reached.

\section{Improvements}

\textbf{Led wavelength.} The spatial resolution of Rasnik is, to first order, proportional to the wavelength of the applied (non-polarised, non-coherent) LED light. In addition, the optimal ChessField size will be smaller for smaller wavelength,
increasing the total contour length in images. The spatial resolution of Rasnik is therefore proportional to
$\lambda^{(1/1.5)}$.
Since powerful UV leds with smaller wavelength are now available, the shortest possible wavelength should be selected. Limits are the transmission of the lens or objective, and the quantum efficiency of the image pixel sensor. The transmission of \textit{reflective} objectives allows a wavelength down to $\SI{200}{nm}$, and image pixel sensors (CCDs) are available with sufficient quantum efficiency for this wavelength. The improvement in spatial resolution is expected to be a factor $3$ when changing the wavelength from $448$ to $\SI{200}{nm}$.

\textbf{Magnification of objective.} Objectives with magnification $100\times$ are available. The correct and stable distance between coded mask and the objective is essential and may cause practical problems.
If UV light is applied, reflection objectives should be used (magnification up to $40\times$). The highest Numerical Aperture (NA) should be chosen in order to minimise image blurring. Changing only the magnification from $20\times$ to $40\times$ should improve the spatial resolution by a factor $2$.

\textbf{Diaphragm, aperture.} For diffraction-limited images projected by a convex singlet, the image blurring Br = $1.22~\lambda/\mathrm{NA}$ (${\mu m}$). For a projecting lens, $\mathrm{NA} = D/b$ where $b$ equals the image distance and $D$ the diameter of the lens (or diaphragm). The lens diameter or diaphragm should chosen as large as possible. For very large aperture, spherical aberration may cause image deformation, but this has no direct effect on the spatial resolution (see section \ref{distortion}).

\textbf{The image pixel sensor.} The spatial resolution of a Rasnik system is inversely proportional to the square root of the surface of the sensor: sensors with a large active surface are advantageous. A multiple of sensors is feasible, provided that the image quality is maintained. Since the \textit{information per unit surface} matters, the pixel pitch is not critical and has a wide optimum. A black-white transition should be covered by minimal 3 pixels: a rule of thumb is that the pixel size should be smaller than $2\times$ the image blurr Br. Small pixels however, with large image data files, require much CPU power.

The spatial resolution of Rasnik systems is determined by the quantum fluctuations of the pixel's content. Increasing the light intensity would improve the spatial resolution, but if the maximum of photo-electrons in a pixel well is exceeded,
the charge will leak away, and the pixel charge content is no longer proportional to the amount of light on that pixel. Increasing the frame rate shortens the exposure time: with the MER-$031$-$860$U$3$M pixel sensor, up to $\SI{860}{fps}$ can be recorded, albeit that substantial computing power is required for the image analysis. The mere increase of frame rate from $\SI{275}{Hz}$ to $\SI{860}{Hz}$ should improve the spatial resolution per $\sqrt{\text{Hz}}$ by a factor $1.7$. Assuming identical well depth, the shorter exposure time will require an inversely higher LED light flux.

\textbf{High full-well capacity (FWC)}. The most important quality factor of state-of-the-art image pixel sensors has been the quantum efficiency which is now approaching the maximum value of 1.
There is now an increasing interest in the development of efficient image pixel sensors that, in addition, can handle high light intensities. In \cite{sensor}, a new sensor is described where a capacitor-per-pixel is applied, increasing the FWC. A further improvement lies in the recording of the pixel current instead of the pixel charge.

\textbf{Data processing.} State-of-the-art image pixel sensors communicate by means of USB$3$. This enables direct access and processing of the image data by any suitable processor. GPUs can (Fourier) process images extremely fast, and multiple GPUs can be controlled by one master processor.
Given commercial interest in machine viewing/learning and in automotive (self-driving), further integration of image sensor and image processing can be expected.

With MicroRas, a spatial resolution was reached of $\SI{7}{pm/\sqrt{Hz}}$. By replacing the Royal Blue LED by a UV led, the Newport M $20\times$ microscope objective by the ThorLabs LMM-$40$X-UVV-$160$ reflective objective, and the MER-$041$-$302$G pixel sensor by the MER-$031$-$860$U$3$M, the spatial resolution could improve by a factor $7$, bringing the performance of Rasnik to the level of $\SI{1}{pm/\sqrt{Hz}}$. With a newly designed ASIC CMOS pixel sensor, a value of $\SI{0.2}{pm/\sqrt{Hz}}$ should be reachable.

\section{Conclusions}\label{conclusions}

Two Rasnik systems were operated while suspended as pendula in vacuum, well-isolated from seismic and acoustical vibrations. The position of the three typical components, namely mask, lens and sensor, were firmly fixed onto a common support. The resulting images were therefore of constant alignment, while the pixel noise is the main variation in the data, revealing the spatial resolution of the system.

The best performance of the SolidRas system is $\SI{110}{pm/\sqrt{Hz}}$, both in linear displacements $x$ and $y$, in good agreement with simulations.
For the resolution of rotation $\theta_z$, an RMS value of $\SI{1.0e-6}{rad/\sqrt{Hz}}$ was measured.

With MicroRas, the image displacement is multiplied with the magnification of the microscope objective  ($20\times$), replacing the lens.
A spatial resolution of $\SI{7}{pm/\sqrt{Hz}}$, both for $x$ and $y$, was reached, again in good agreement with simulations. Given the diffraction-limited images, the performance of Rasnik is limited by the quantum fluctuations of the light, absorbed by the pixels.

The demonstrated performance of MicroRas of $\SI{7}{pm/\sqrt{Hz}}$ can be improved by using a UV led, by replacing the microscope objective by a reflective objective, and using a faster version of the image pixel sensor: a value of $\SI{1}{pm/\sqrt{Hz}}$ should be within reach, with off-the-shelf components. By developing a special image pixel sensor capable to handle a higher light flux, the spatial resolution could be further improved by a substantial factor.

The following features make the Rasnik systems compatible or advantageous with respect to other displacement sensors:

\begin{itemize}
	\item Rasnik measures a $2$D displacement in the plane of a planar object (mask) with respect to the optical axis, perpendicular
	to the mask plane. The optical axis is defined by the fixed-together lens and optical image sensor unit
	\item systematical linearity error $\SI{50}{pm}$ or smaller over arbitrarily large dynamic range, in $x$ and $y$
	\item only one precision element: the coded mask, available at low cost thanks to MEMS and IC industry
	\item no cross coupling between output parameters $x$, $y$, S and $\theta_z$; perpendicularity between $x$ and $y$ only defined by the mask
	\item no calibration required
	\item straight, analog/digital linear measurement system: no ‘lock’ control and feedback systems required
	\item arbitrarily large range of measurement (determined by mask size)
	\item extremely low Flicker noise
	\item no special electronics required: a system can be composed of only commercially available CMOS pixel image sensors,  CPUs, GPUs and USB or Ethernet$3$ networking, and COTS optical components
	\item (very) low cost: $300$ - $3000$ € excl. network and CPUs
	\item light pressure onto the mask acts perpendicular to the directions of interest
	\item system can operate in air, or in vacuum, and in cryogenic environments if case-specific precautions are taken.
\end{itemize}

The proven spatial resolution in $x$ and $y$ of $\SI{7}{pm/\sqrt{Hz}}$ is not as good as can be obtained with interferometers
($\SI{4}{fm/\sqrt{Hz}}$, see ~\cite{Heijningen_2020}) and SQUIDs ($0.1$ to $\SI{3}{fm/\sqrt{Hz}}$, see~\cite{sato}). Another disadvantage of Rasnik is that it requires substantial CPU power, albeit that this is the focus of future (machine viewing, LADAR, self-driving cars) developments.

~\newline
\noindent\fbox{%
    \parbox{0.975\textwidth}{%
        \textbf{Because of its low cost and simplicity, Rasnik could be well-applied in a large area network
        of multiple seismic sensors, and in deformation-sensing instrumentation in general.}
    }%
}

\section*{Acknowledgements}
We would like to thank Wim Gotink for his electro-mechanical assistance, Martijn van Overbeek, Berend Munneke and Krista de Roo for their assistance to the vacuum tank facility.
We are grateful for the high-precision mechanical parts made by Oscar van Petten.

\appendix
\section{Appendix: Applications}

\begin{figure}[htb]
\centering
\includegraphics[width=15cm]{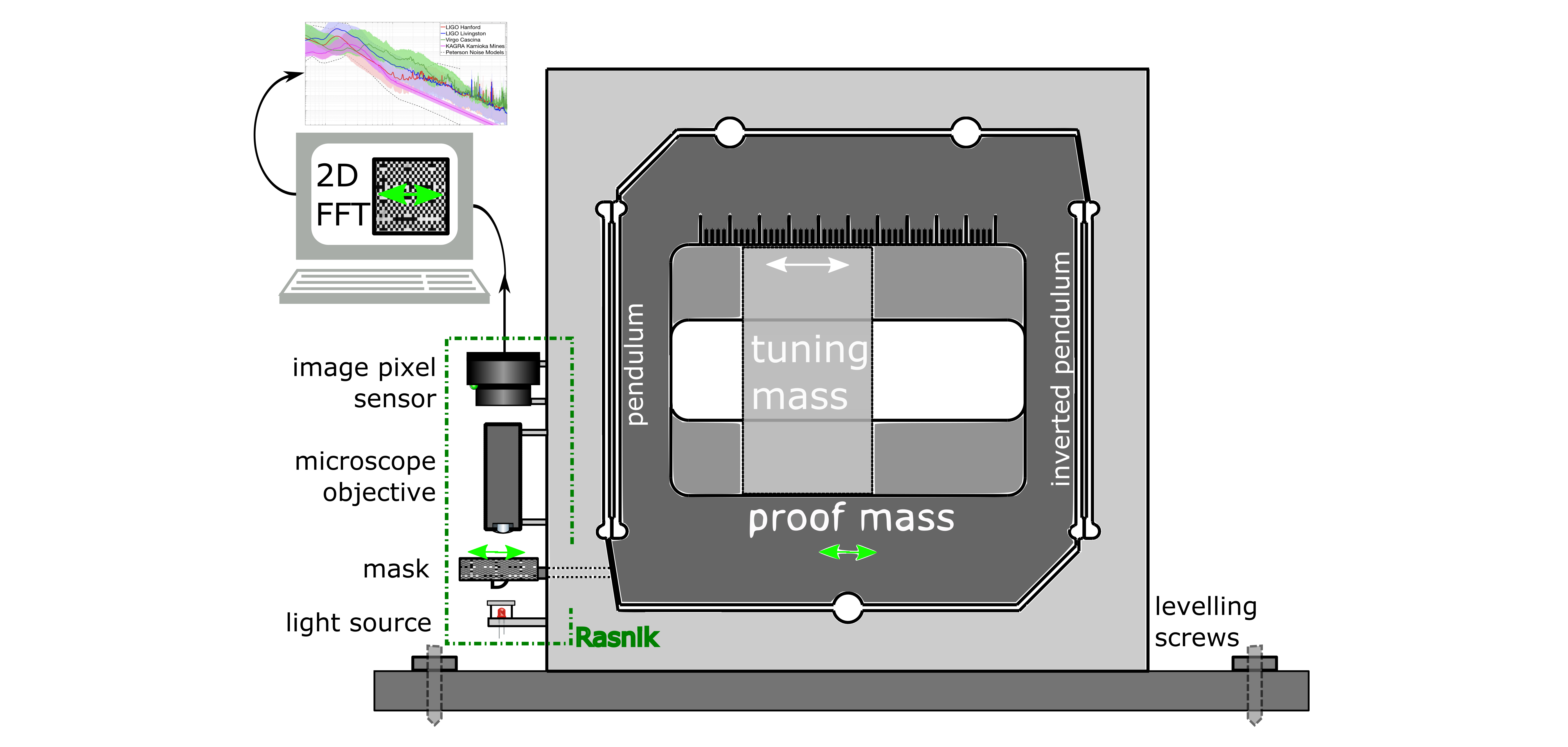}
\caption{Accelerometer with a Rasnik readout:the position of the proof mass with respect to the base frame is probed by a Rasnik system.}
\label{fig:joris}
\end{figure}

\textbf{Seismic sensor: accelerometer}. Rasnik could be applied in instruments for registering seismic motions in general~\cite{watt}. A practical application is shown in Fig~\ref{fig:joris}. The proof mass is suspended in a Watt’s linkage configuration which allows for low frequency tuning and thus increased mechanical sensitivity. The position of the proof mass with respect to the ground-attached frame is recorded by one or more Rasnik systems: each system measures two degrees of displacement freedom. With one Rasnik system, both the horizontal (x and y) acceleration is measured, in absence of cross-coupling between the two. With three Rasnik systems, a 6D inertial sensor can be realised. The Rasnik systems are operational as soon as the proof mass is released, and no feedback systems are required.
With the high-precision mask as unique reference, calibration is not needed, omitting calibration hardware. Given the large (mm) range of operation, the systems will continue to operate after high-amplitude events like earthquakes or storms.

\textbf{The Moon as antenna for gravitational waves}. In reference~\cite{Harms}, a network of seismic sensors is proposed, using the Moon as object of which seismic waves, deformations and vibrations are to be recorded. The relevant information is embedded in the phase \textit{difference} in the seismic signals from multiple sensors distributed over a certain area. By placing Rasnik systems on the Moon's surface spanning a large distance (10 - 50 km) between the light source and image sensor, the relevant differential displacements can be recorded~\cite{slac}. In absence of feedback systems, the Lunar Rasnik system would be operational immediately after the installation of the three components including the power- and data systems. \label{appendix}


\end{document}